\documentclass[final,3p,times]{elsarticle}

\usepackage{geometry}
\usepackage{amsmath}
\usepackage{physics}
\usepackage{MnSymbol}
\usepackage{graphicx}
\usepackage{geometry}
\usepackage[table,svgnames]{xcolor}
\definecolor{yasi}{HTML}{8554ed}
\definecolor{sabz}{rgb}{0.5, 1.0, 0.83}
\definecolor{arghavan}{rgb}{0.74, 0.2, 0.64}
\usepackage{booktabs}
\newcommand{\cellg}{\cellcolor{yasi}}
\newcommand{\cellb}{\cellcolor{sabz}}
\newcommand{\cellr}{\cellcolor{arghavan}}

\usepackage{layout}

\begin{document}

\begin{frontmatter}

\title{Entanglement dynamics and fractional quantum state transport in the spin-$\frac{1}{2}$ triangular plaquette}

\author{M. Motamedifar}

\address{Department of Physics, Shahid Bahonar University of Kerman, Kerman, Iran}

\begin{abstract}
The dramatic growth of research areas within the province of quantum state transmission (QST) is rapidly accelerating. An important insight to understand the process of QST can be fulfilled by considering the dynamical behavior of its entanglement content.  One well-established approach to continuously transfer quantum states is utilizing spin structures. Here, from the view of entanglement propagation,  we disclose the signature of fractional QST possibilities. In the present work, we proposed a form of spin-$\frac{1}{2}$ triangular plaquette whose Hamiltonian entails the spin-orbit coupling on the rungs and exchange interaction over the legs. The feature of such a system is that the time instants of QST emerge in a discrete fashion, thereby the values of exchange interactions associated with these moments behave fractionally. Importantly, it is found that for special values of magnetic interaction, QST has singularity i.e., the entanglement propagation is forbidden.  In addition, the finite-size nature of this system makes it possible for us to read the nexus between time crystallinity and symmetry breaking. The development of our knowledge about time crystalline symmetry and its breaking helps us to understand the defined concept and fundamental physics of this phenomenon.
\end{abstract}


\end{frontmatter}


\section{Introduction}
\label{intro}

Quantum state transfer (QST) is one of the essential requirements for the processing of quantum information.  Depending on the available technology, alternative methods have been designed to perform the propagation of quantum states.
Photons as flying qubits have a well-established quantum communication ability~\cite{chen2015expected}.
On the other hand, spins as the stationary solid-state qubits play the role of information carriers that can be hosted by chain structure as Bose suggested~\cite{bose2003quantum}. He propounded that in the course of time, a chain of coupled spins would develop a quantum state.
For short-range quantum communication in which qubits must be close enough to each other, quantum spin chains are attractive candidates. On the other hand, photons in such systems suffer several phenomena arising from light-matter interaction which renders them useless for practical ideas.

Following Bose's seminal suggestion, a lot of effort has been put into realizing such an scenario in quantum computing technology. However, it was found that the perfect QST can be reached through a four-spin chain~\cite{bose2003quantum, wang2001entanglement, feng2007entanglement,motamedifar2017dynamical,motamedi2019exact}, much recent research has concentrated on achieving high fidelity of the state traveling across many-qubit systems~\cite{wang2008quantum, lyakhov2007role, boness2010doubly, oh2011heisenberg, giorgi2013quantum, lorenzo2015transfer}. The role of magnetic exchange interaction between the spins, which is of the Heisenberg-type, is being explored in many of these systems~\cite{subrahmanyam2004entanglement,apollaro2015many,vieira2019robust,apollaro2019spin}. In fact, the Heisenberg model is at the core of magnetic models in low dimensional materials. In addition, the role of another magnetic interaction stemming from spin-orbit mechanism i.e., Dzyaloshinskii-Moryia (DM) coupling in QST has been recently studied~\cite{shi2017robust, mahmoudi2019effects}. Such an interaction triggers various spin patterns in different quantum phases~\cite{messio2017chiral,lee2018magnonic,farajolla} as well as entails problems such as spintronics~\cite{kato2019current}, multiferroics~\cite{stagraczynski2017many} and topological magnetic orders~\cite{yokouchi2017electrical}, just to name a few.

In addition to being hosted by chains, spin qubits can also be accommodated in other configurations.
For instance, we can address the triangular plaquette which includes two legs and three rungs.
In Ref.\cite{motamedi2019exact} it is assumed that the Heisenberg interaction is designated on the rungs of the plaquette, in addition to DM coupling over the legs.
Exploring entanglement evolution shows that through the entanglement oscillation, system is able to transmit quantum state continuously.
Consequently, it is found that the appearance of quantum $W$ states due to the dynamics of the system is of interest.

The present work, as set forth later in more detail, indicates that this process takes place in a fractional fashion, provided that in comparison with the situation of Ref.~\cite{motamedi2019exact}, the plaquette's links become exchanged to host magnetic interactions. This means that the QST will arise only for discrete values of Heisenberg coupling over the legs, for a fixed value of DM interaction on rungs.
Nevertheless, the cost of such a feature would however be to sacrifice the quantum $W$ states.

In this situation, the QST is possible only for fractional values of magnetic interaction over the legs of the plaquette. The existence of fractional magnetic coupling for licensing QST is not a generic status. For example, the necessity of particular Heisenberg interaction in a perfect quantum-state transition has already been discussed in Ref.\cite{christandl2004perfect}. 
In addition, we encountered situations where the entanglement is locked at one part of the system. This is where, the particular values of Heisenberg interaction induces singularity in QST. From thermal states perspectives, allowed and forbidden bipartite correlations has been argued in Ref.~\cite{guha2019allowed}.

Time crystallinity is a phenomenon which first put forward by Wilczek~\cite{wilczek2012quantum}‌. A closed System' Hamiltonian induces an oscillatory behavior in expectation values of a physical observable via the time evolution of the system, thereby that time translation symmetry is broken spontaneously. This phenomenon is called the breaking of time crystalline whose name is borrowed of a similar condition that occurs in crystals when the spatial translation symmetry breaks due to the periodic arrangement of atoms~\cite{oberreiter2020stochastic}.  In the present survey, entanglement oscillation that is a direct result of quantum correlations' behavior reveals how crystal time symmetry is broken.

For dealing with the above situations in details, the outline of the article is organized as follows: In Sect.\ref{sec:model} we implement the model under consideration and explain how the initial state progresses over time intervals.  
The start of Sect.\ref{sec:Introducing tools} is dedicated to provide relevant tools and concepts used in this work. The rest of this section devoted to present the analytical results. Finally Sect.\ref{sec:concl} summarizes conclusions and outlooks.

\section{The model and state evolution}
\label{sec:model}
Here, we consider a triangular plaquette in which spin$-\frac{1}{2}$ particles interact on the rungs with DM coupling while interacting on the legs through isotropic Heisenberg Hamiltonian.
Under the periodic boundary condition on rungs, the system is described by the Hamiltonian:
\begin{eqnarray}
\hat{H}=\vec{D}
\boldsymbol{\cdot}\sum_{\langle i,j\rangle}\hat{\vec{S}}_{i}\times\hat{\vec{S}}_{j}+J\sum_{\llangle i',j'\rrangle}(\hat{S}^{x}_{i'}\hat{S}^{x}_{j'}+\hat{S}^{y}_{i'}\hat{S}^{y}_{j'}+\hat{S}^{z}_{i'}\hat{S}^{z}_{j'}),
\label{eq:Ham}
\end{eqnarray}
where $\vec{D}=D\hat{z}$ on the right side of Eq.(\ref{eq:Ham}) shows the strength of DM interaction on the rungs ($\langle i,j\rangle$). Additionally, $J$ in the second term of Eq.(\ref{eq:Ham}) stands for the strength of Heisenberg coupling over the legs ($\llangle i',j'\rrangle$).

Considering the time-independent Hamiltonian which undertakes the dynamical behavior of such a closed system, we begin with Shr\"{o}dinger equation:
\begin{eqnarray}
\partial_t \ket{\psi} = -\frac{i \mathbf{H}}{\hbar}\ket{\psi} \rightarrow \ket{\psi(t)} = e^{\frac{-i \mathbf{H} t}{\hbar}} \ket{\psi(0)}, \label{eq:unit_evo}
\end{eqnarray}
Taking into account the system's Hamiltonian (\ref{eq:Ham}), $t$ is normalized to $\hbar/D$ and $J$ is normalized to $D$. No loss of generality we take $\hbar\equiv 1$, and consequently, $t$ and $J$ converts to dimensionless parameters.

The initial state of the system ($\ket{\psi(0)}$) is given by quantum state $\frac{1}{\sqrt{2}}(\ket{10} _{12} + \ket{01}_ {12})\ket{00}_{34}$, in which two first qubits $(1,2)$ are entangled via a Bell state and two last qubits are in a separable state. In such a state, $\ket{0}$ indicates the spin down state (i.e., spin oriented along $-z$ orientation), while $\ket{1}$ denotes the reversed direction. Upon the mentioned initial state, the general form of $|\psi(t)\rangle$ obeys:
\begin{eqnarray}
|\psi(t)\rangle=\sum_{i=1}^{16}\sum_{j=1}^{16}a_i b_j e^{-i E_i t}|\alpha_j\rangle.
\label{eq:unitapli2}
\end{eqnarray}
where $|\alpha_j\rangle$ shows an element in the standard bases i.e., $\{|1111\rangle, |1110\rangle,...,|0000\rangle\}$. Furthermore, $a_i=\bra{E_i}\psi(0)\rangle$ and  $b_j=\bra{\alpha_j}E_i\rangle$ where $\ket{E_i}$ is i-th energy eigenvector according to energy $E_i$.

After a straightforward calculation, we note that all of the coefficients in Eq.~(\ref{eq:unitapli2}) are null except those which correspond to the single-particle states. In such states a particle is in the $+z$ direction of the spin space and the rest are in the direction of $-z$.
As a result, the system's evolved quantum state is as follows:
\begin{eqnarray}
\ket{\psi(t)}=\eta |0001\rangle+\xi |0010\rangle+\alpha |0100\rangle+ \beta |1000\rangle.
\label{eq:kett1}
\end{eqnarray}
where
\begin{eqnarray}
\eta&=&\frac{1}{2\sqrt{2}}\left[\cos \left(\frac{J t}{2}\right) (\sin (D t)-\cos (D t)+1)-i \sin \left(\frac{J t}{2}\right) (-\sin (D t)+\cos (D t)+1)\right]\nonumber\\
\xi&=&\frac{1}{2\sqrt{2}}\left[\cos \left(\frac{J t}{2}\right) (-\sin (D t)-\cos (D t)+1)-i \sin \left(\frac{J t}{2}\right) (\sin (D t)+\cos (D t)+1)\right]\nonumber\\
\alpha &=&\frac{1}{2\sqrt{2}}\left[\cos \left(\frac{J t}{2}\right) (-\sin (D t)+\cos (D t)+1)+i \sin \left(\frac{J t}{2}\right) (-\sin (D t)+\cos (D t)-1)\right]\nonumber\\
\beta &=&\frac{1}{2\sqrt{2}}\left[\cos \left(\frac{J t}{2}\right) (\sin (D t)+\cos (D t)+1)+i \sin \left(\frac{J t}{2}\right) (\sin (D t)+\cos (D t)-1)\right]\nonumber
\end{eqnarray}
At each time instant, a linear combination of one-particle states emerges in Eq.(\ref{eq:kett1}). The measure of the entanglement between different parts of the system will be discussed in what follows, using this equation.
\section{Introducing tools and presenting results}
\label{sec:Introducing tools}
As proposed by Wootters et al.\cite{hill1997entanglement,wootters1998entanglement}, the concurrence is a perfect quantifier for the measure of entanglement between the m-th qubit of a quantum system with n-th one of that. Along such a suggestion, the starting point is to provide the reduced density matrix which is obtained by tracing out all qubits except two target ones $m$ and $n$ ($\rho_m^n=Tr_{m,n}(\rho)$). The second step is to construct the matrix $R=\rho_m^n.\hat{\tau}.\rho_m^{n*}.\hat{\tau}$ using $\hat{\tau}=\sigma_y\otimes\sigma_y$ where $\sigma_y$ and $\rho_m^{n*}$ stand for the Pauli-y-operator and the complex conjugate of $\rho_m^{n}$ respectively. The relation between ($\mathcal{C}^m_n$) and the eigenvalues of matrix $R$ is defined by $\mathcal{C}^m_n=max\{2\gamma_1-\sum_i \gamma_i,0\}$ in which $\gamma_1$ is the largest eigenvalue.
By definition, $0 <\mathcal{C}^m_n< 1$ shows that two qubits are partly entangled. Whereas $\mathcal{C}^m_n=1$ corresponds to a state of maximal entanglement, $\mathcal{C}^m_n=0$ describes a state of total disentanglement between two qubits. Making use of Eq.~(\ref{eq:kett1}), our calculation reveals that concurrences for any pairs of the system in question, are supplied by Eqs.~(\ref{eq:concurencess}):
\[
\begin{split}
\mathcal{C}^2_1=\frac{1}{32}\Bigl(
& 2 \cos \bigl(t (J-3)\bigr)
+2 \cos \bigl(2 t (J-1)\bigr)
+12 \cos (t) \cos (t J)
\\
&+2 \cos \bigl(2 t (J+1)\bigr)
+2 \cos \bigl(t (J+3)\bigr)
+4 \cos (2 t)+\cos (4 t)+7
\Bigr)\nonumber\\
\mathcal{C}^4_3=\frac{1}{32}\Bigl(
& -2 \cos \bigl(t (J-3)\bigr)
+2 \cos \bigl(2 t (J-1)\bigr)
-12 \cos (t) \cos (t J)
\\
&+2 \cos \bigl(2 t (J+1)\bigr)
-2 \cos \bigl(t (J+3)\bigr)
+4 \cos (2 t)+\cos (4 t)+7
\Bigr)\nonumber\\
\mathcal{C}^3_1=\frac{1}{32}\Bigl(
& -2 \sin \bigl(2t (J+1)\bigr)
-2 \sin \bigl(2 t (1-J)\bigr)
-4 \sin (2t)-\cos(4 t)+5
\Bigr)
\end{split}
\label{eq:concurencess}
\]

\begin{figure*}[!t]
	\centering
	\begin{tabular}[b]{c}
		\includegraphics[width=.45\linewidth]{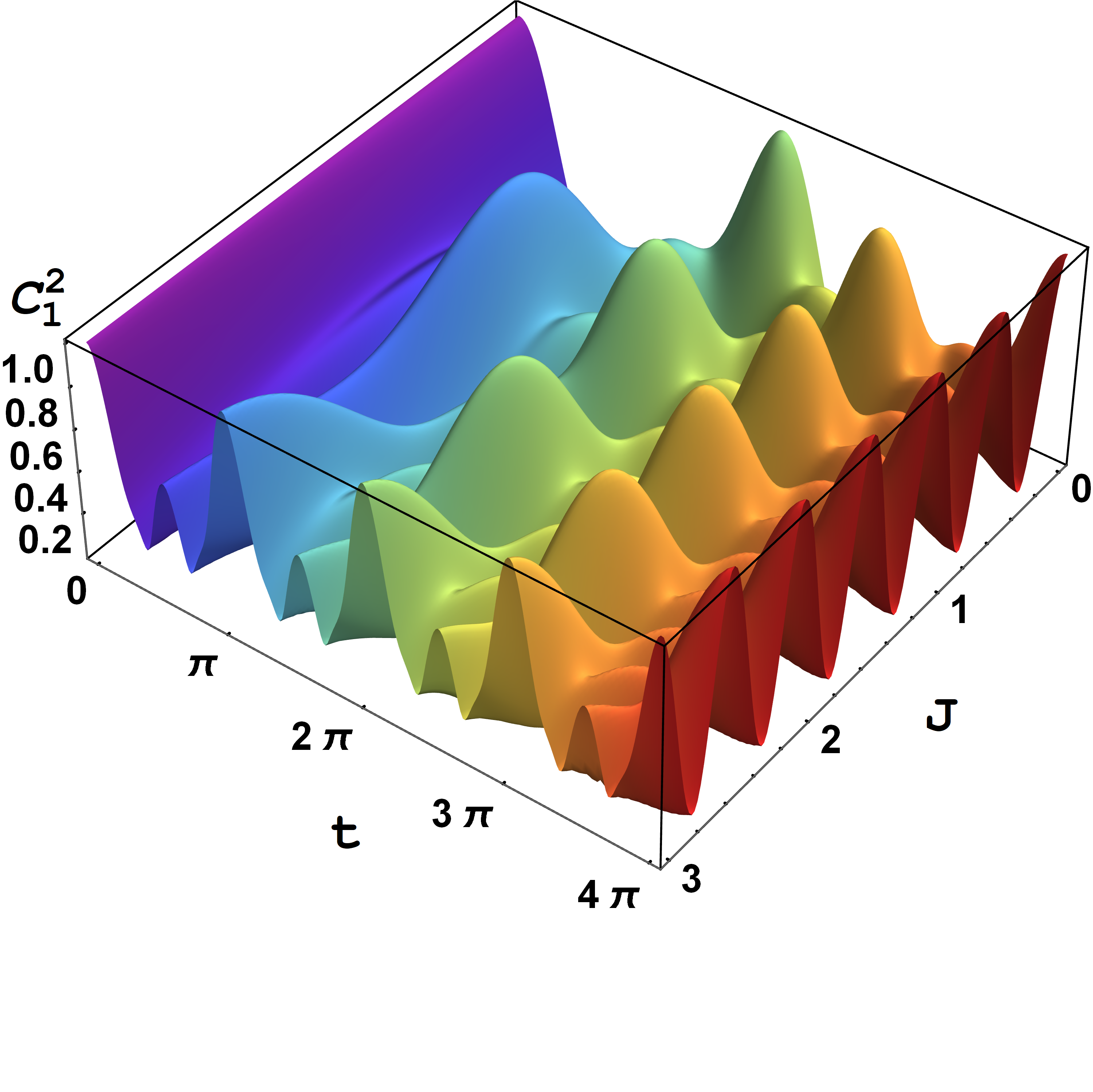}\\
		\small (a)
	\end{tabular}
	\begin{tabular}[b]{c}
		\includegraphics[width=.45\linewidth]{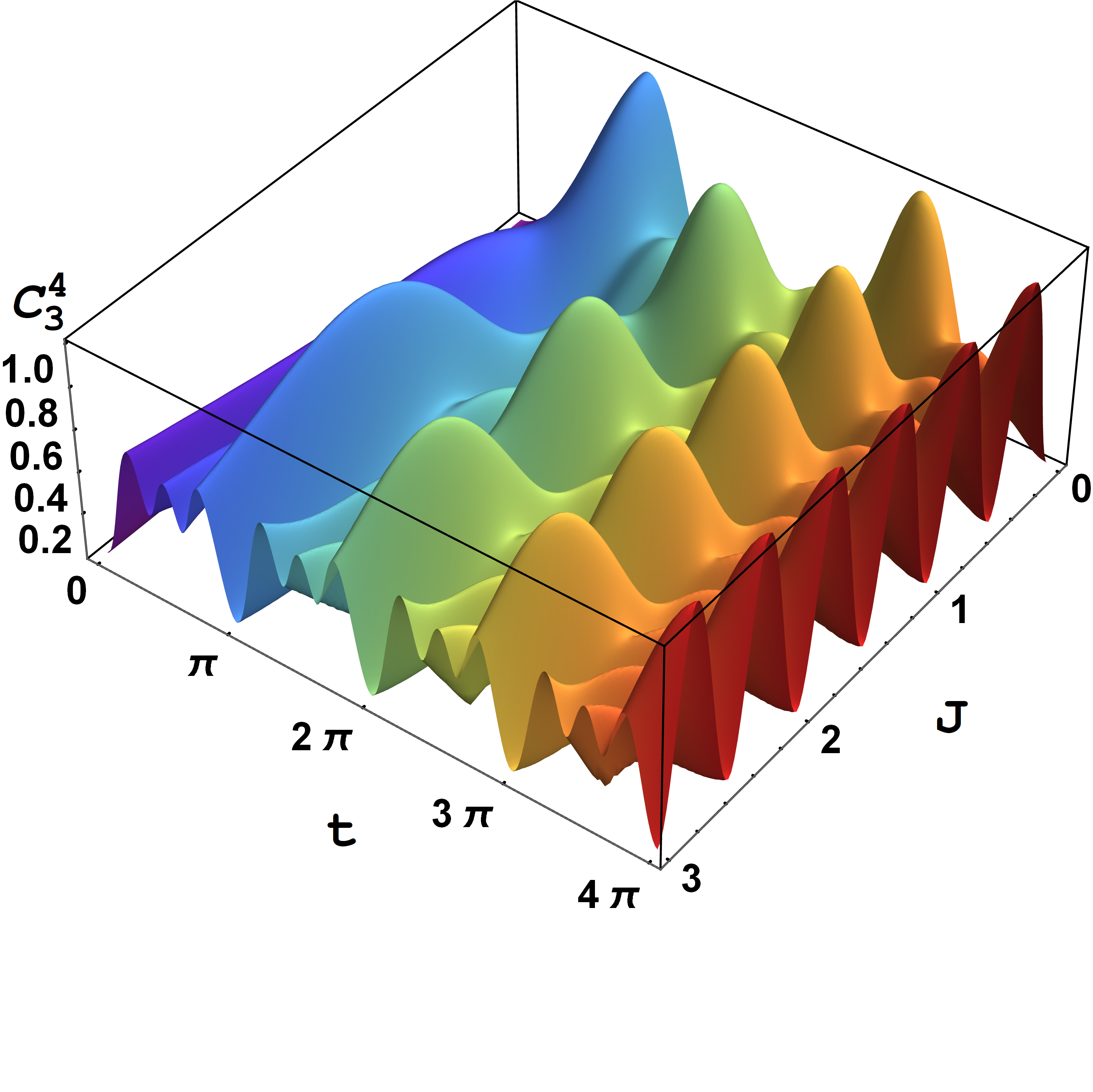}\\
		\small (b)
	\end{tabular} 
	\begin{tabular}[b]{c}
		\includegraphics[width=.45\linewidth]{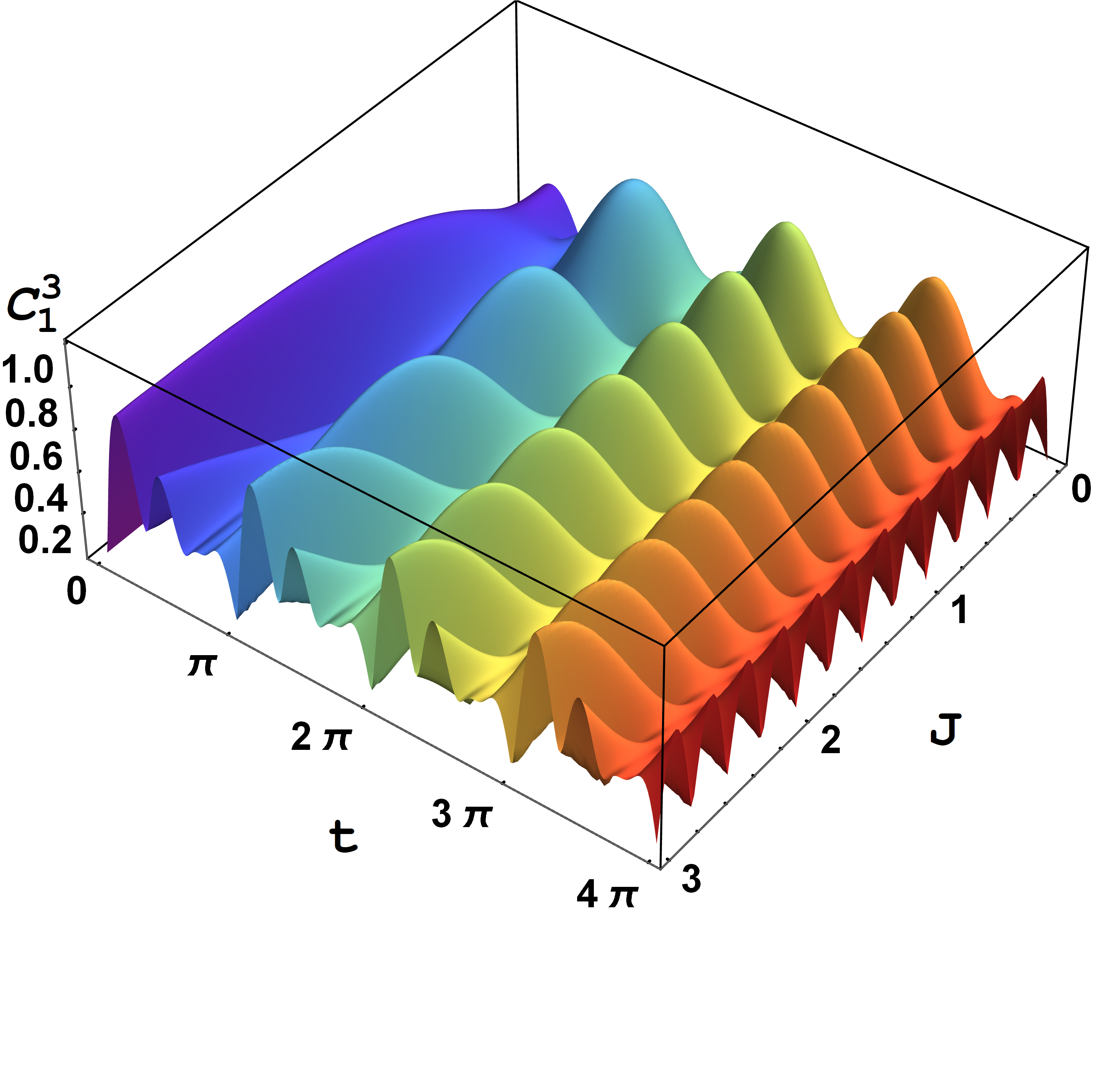}\\
		\small (c)
	\end{tabular} 
	\caption{\footnotesize{ The illustration of concurrence's behavior versus $J$ and $t$ shared on (a) the pair of ($1,2$), (b) the pair of ($3,4$) and (c) the pair of ($1,3$) that is representative of other pair i.e., ($2,4$).}}
	\label{fig:Fig13D}
\end{figure*}

\begin{figure*}[t]
	\centering
	\begin{tabular}[b]{c}
		\includegraphics[width=.50\linewidth]{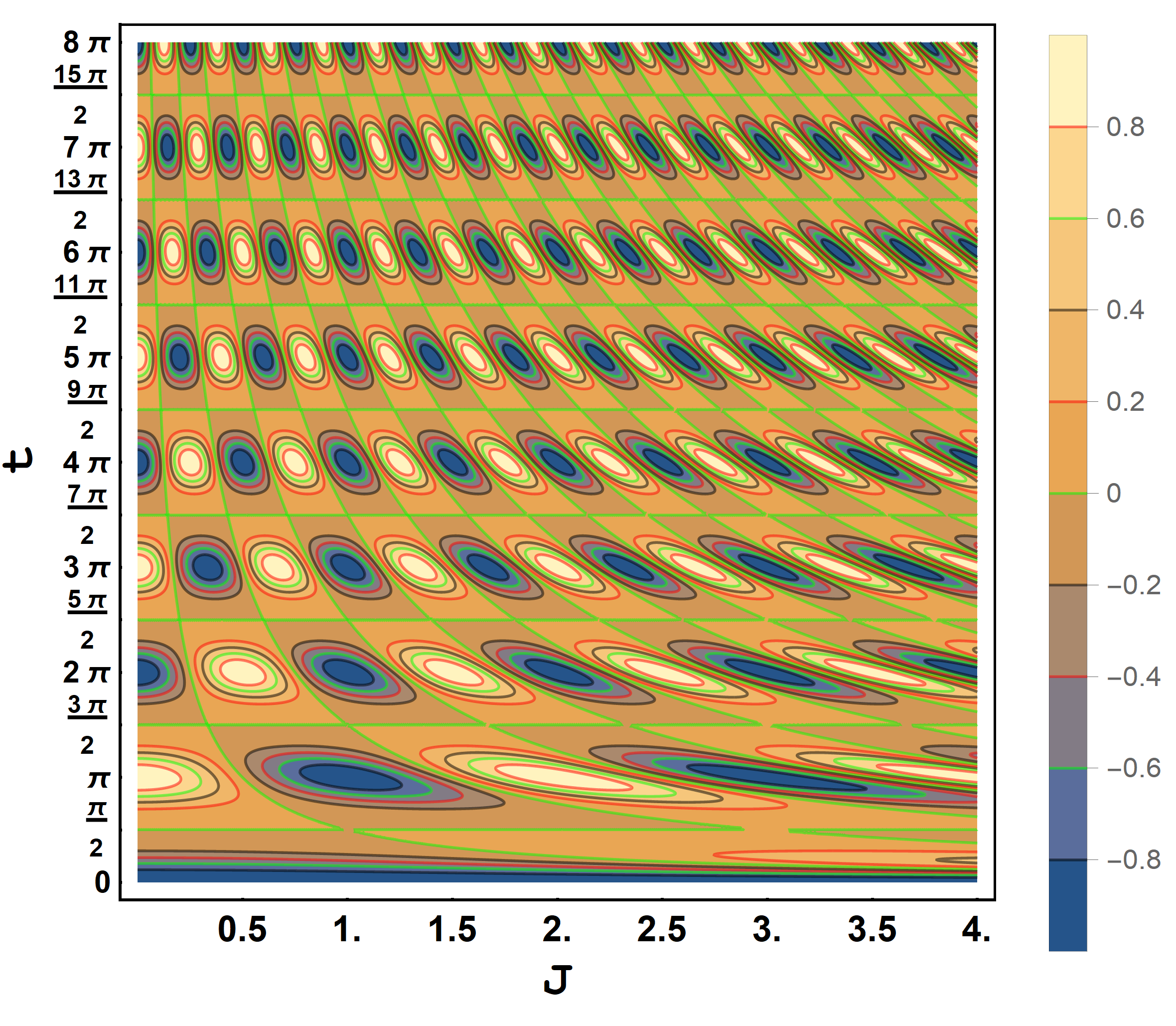}\\
		\small (a)
	\end{tabular}
	\begin{tabular}[b]{c}
		\includegraphics[width=.43\linewidth]{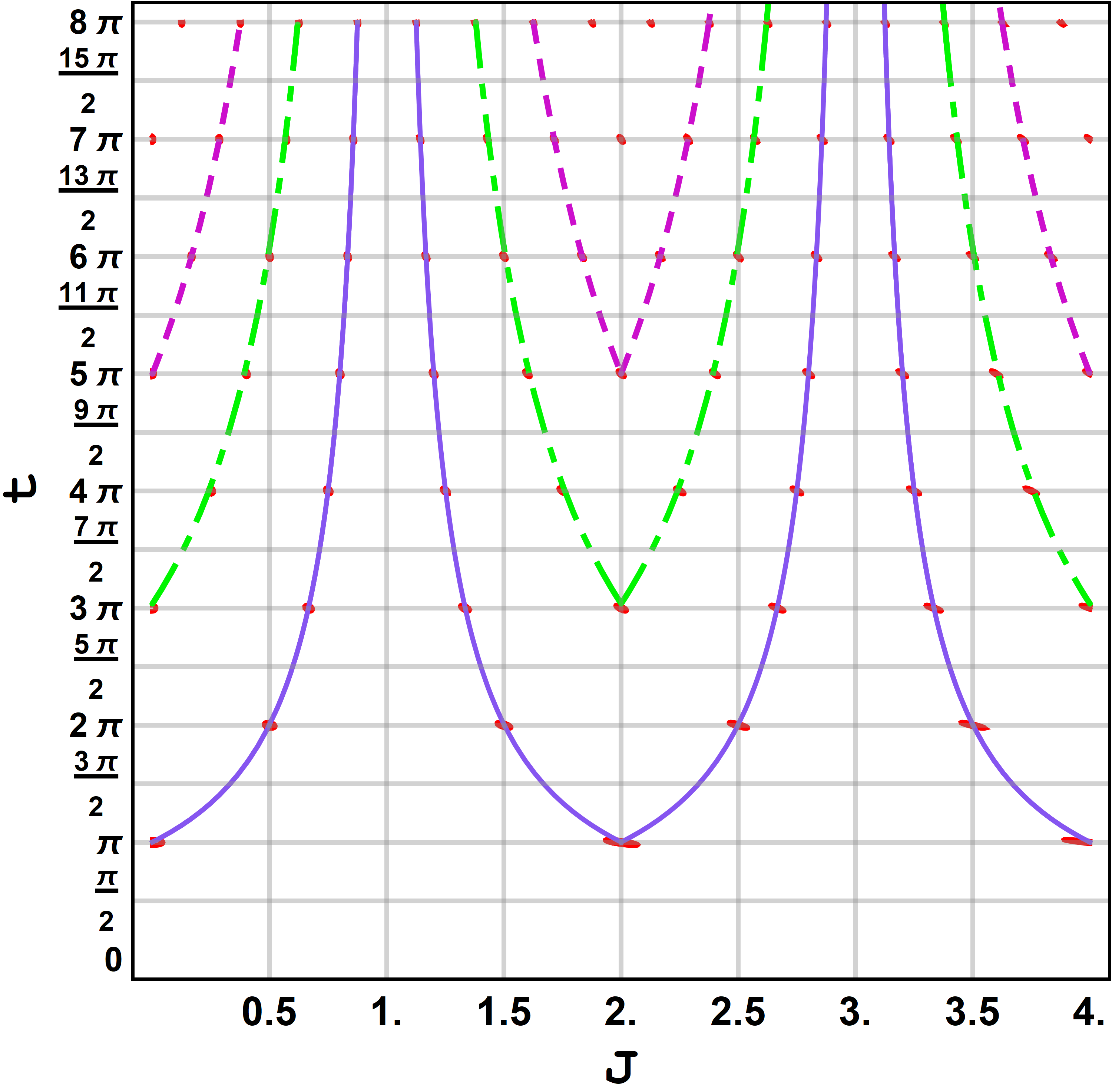}\\
		\small (b)
	\end{tabular} 
	\caption{ \footnotesize{(a) Depiction of $\mathcal{G}^{34}_{12}(t,J)$ to recognize the points of QST. The center of bright points indicate QSTs, (b) The red dots are corresponded to the roots of $\mathcal{G}^{34}_{12}(t,J)-1=0$ and according to the center of bright points of the left plot. Such roots follow from determined sequences over individual curves. Each curve obeys a distinct sequence  formula that presented in Table.\ref{tab:tab} }}
	\label{fig:diff}
\end{figure*}

Figures \ref{fig:Fig13D}(a) and \ref{fig:Fig13D}(b) describe the concurrence's dynamical behavior  between the first and last pairs respectively. Moreover, Fig.\ref{fig:Fig13D}(c) is devoted to $\mathcal{C}^3_1$. All panels show the oscillation behavior versus time and Heisenberg strength, however, it is difficult to see conditions for QST in which $\mathcal{C}_1^{2}=0$ and $\mathcal{C}_3^4=1$.
Then, as we are looking for the time instants when QST happen, we should calculate the difference (gap) between $\mathcal{C}^2_1$ and $\mathcal{C}^4_3$ which brings us to Eq.(\ref{eq:tafaazol}):
\begin{eqnarray}
\mathcal{G}^{34}_{12}(t,J)=\mathcal{C}^4_3-\mathcal{C}^2_1=\frac{1}{8} \Bigl(-\cos\bigl(t ( J-3)\bigr) - 6 \cos(t) \cos(t J) - \cos\bigl(t (3 + J)\bigr)\Bigr),
\label{eq:tafaazol}
\end{eqnarray}
which is illustrated in Fig.\ref{fig:diff}(a). Consequently, we can conclude that when such a gap equates to unity, the QST appears. As it can be seen, the presence of alternating bright points in this plot indicates that QST does not happen continuously. These points take the stand that QSTs are arose. Focusing on such a panel, it is observed that the
centers of the bright spots are precisely in front of the integral multiple of $\pi$ i.e., time instants of QSTs equate with $m\pi$ ($\mathcal{T}_{tr}=m\pi$) provided that $m=\{1,2,3,...\}$.

\begin{table}
	$\begin{array}{ l *{9}{c} }
	\toprule
	m & \multicolumn{2}{c}{J^*(m)}     & t^*(m)
	& \multicolumn{2}{c}{J^{**}(m)}  & t^{**}(m)
	& \multicolumn{2}{c}{J^{***}(m)} & t^{***}(m) \\
	\cmidrule(lr){2-3} \cmidrule(lr){5-6} \cmidrule(lr){8-9}
	& 1-1/m & 1+1/m & m\pi
	& 1-3/m & 1+3/m & m\pi
	& 1-5/m & 1+5/m & m\pi \\
	\midrule
	\cellg 1 & \cellg 0   & \cellg 2   &  \cellg \pi \\
	2 & 1/2 & 3/2 & 2\pi \\
	\cellg 3 & \cellg 2/3 & \cellg 4/3 & \cellg 3\pi & \cellb 0   & \cellb 2   & \cellb 3\pi \\
	4 & 3/4 & 5/4 & 4\pi & 1/4 & 7/4 & 4\pi \\
	\cellg 5 & \cellg 4/5 & \cellg 6/5 & \cellg 5\pi & \cellb 2/5 & \cellb 8/5 & \cellb 5\pi & \cellr 0 & \cellr 2 & \cellr 5\pi \\
	6 & 5/6 & 7/6 & 6\pi & 3/6 & 9/6 & 6\pi & 1/6 & 11/6 & 6\pi \\
	\cellg 7 & \cellg 6/7 & \cellg 8/7 & \cellg 7\pi & \cellb 4/7 & \cellb 10/7 & \cellb 7\pi & \cellr 2/7 & \cellr 12/7 & \cellr 7\pi \\
	\bottomrule
	\end{array}$
	\caption{\footnotesize{Fractional value of Heisenberg interaction for interval $J\in \left[0,2\right]$ corresponded to QST happening. The color of any column is according to the color of any curve in Fig.~\ref{fig:Fig13D}}}
	\label{tab:tab}
\end{table}
Furthermore, red spots in Fig.\ref{fig:diff}(b)  coincide with centers of bright points in Fig.\ref{fig:diff}(a). There are different curves in Fig.\ref{fig:diff}(b) that can pass over the red dots. As mentioned before $\mathcal{T}_{tr}=m\pi$, whereby the values of $J$ which satisfy $\mathcal{G}^{34}_{12}(t=\mathcal{T}_{tr}, J)-1=0$ have to be discrete. For example, regard to solid curve in Fig.\ref{fig:diff}(b), we have $J=\{0,\frac{1}{2},\frac{2}{3},\frac{3}{4},...\}$ in ascending order for $0\le J<1$. On the other side, for $1<J\le 2$ we have $J=\{2,\frac{3}{2},\frac{4}{3},\frac{5}{4},...\}$ in descending order. Without any generality we restrict ourselves to $J\in \left[0,2\right]$, because every thing is repeated after this period. For such an interval the sequence  formula of allowed $J$ values for such a curve as it presented in Table.\ref{tab:tab} obeys from $J^{*}=1-1/m$ and $J^{*}=1+1/m$ when $m=\{1,2,3,...\}$.

For the second curve (dash-dot one) in Fig.\ref{fig:diff}(b) it is found that for $\mathcal{T}_{tr}=m\pi$ with $m=\{3,4,5,...\}$ the Heisenberg strength for $0\le J<1$ must be chosen from set of sequence $J=\{0,\frac{1}{4},\frac{2}{5},\frac{3}{6},...\}$ and for $1<J\le 2$, they must be selected from $J=\{2,\frac{7}{4},\frac{8}{5},\frac{9}{6},...\}$. The sequence  formula of special $J$ values for such a curve as it presented in Table.\ref{tab:tab} obeys from $J^{**}=1-3/m$ and $J^{**}=1+3/m$ when $m=\{3,4,5,...\}$.
Subsequently, a similar argument holds for the dashed curve in Fig.\ref{fig:diff}(b) in which $m\ge 5$ and we have $J^{***}=1-5/m$ and $J^{***}=1+5/m$. However there are other curves for the rest of $J$ values, we finish this argument here.  

Interestingly, we notice here that for some particular situations i.e., $J=1,3,...$, the entanglement transition is forbidden. It means that for such values of $J$, the equation $\mathcal{G}^{34}_{12}(\mathcal{T}_{tr}, J)-1=0$ does not have any root. As though, the last pairs cannot received the sent state encoded by first pairs. 

\begin{figure}
	\centering
	\includegraphics[width=0.5\linewidth]{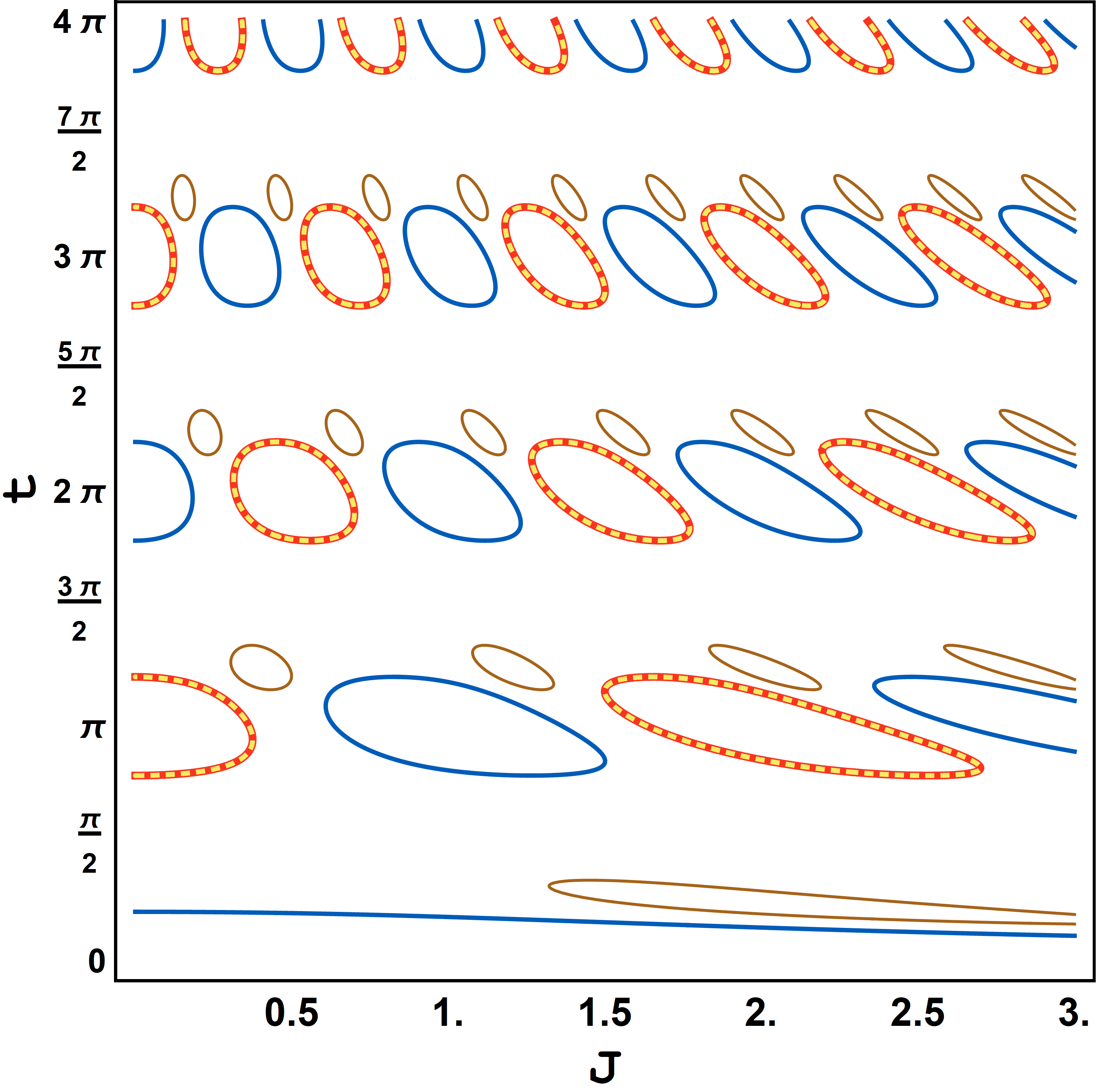}
	\caption{\footnotesize{The illustration of concurrences' contours that all are equal to $0.5$. The blue (thick-solid) curve is related to $\mathcal{C}^2_1$, dashed one in corresponded to $\mathcal{C}^4_3$ and brown (thin-solid) curve is according to $\mathcal{C}^3_1$. Since there is no any cross between curves, quantum $W$ states are not emerged}. }
	\label{fig:aslwstatestime}
\end{figure}

To check on whether the quantum states of $W$ type can be derived from the evolution of the system we follow the discussion. Since all pairs have generally the same concurrences in quantum $W$ states equal to $2/N$, where $N$ is the particle number of the system, then we conclude that all concurrences would be equal to $0.5$ for four qubits if those states occur. For this reason we present the Fig.\ref{fig:aslwstatestime}. In other words, a plane that cuts the axis of concurrences of Fig.~\ref{fig:Fig13D} in half. Figure~.\ref{fig:aslwstatestime} includes three types of curves. As shown this panel, there are no crossing points that can be shared between these three curves consequently it is a witness that
$W$ states do not exist.

\section{Conclusion}
\label{sec:concl}
The field of quantum state propagation has been greatly enhanced by a range of recent developments in both quantum information science and quantum magnetism. The synergy between quantum information community and world of quantum magnetism offers a wealth of applications such as utilizing spin structures in entanglement propagation.  The current study deals with the spin-$\frac{1}{2}$ triangular plaquette in which Heisenberg interaction coexists with DM interplay. The role of present survey is not only to account for rendering fractional QST behind entanglement propagation, it also provides a situation of entanglement locking. We have seen how fractional values of Heisenberg magnetic interaction over the legs of plaquette lead distinct instants of QST. In order to detect such values of interaction we made use of the gap between the concurrences of the system's last and first pairs. Equalizing this gap with unity ensures that the entanglement is propagated to the last pairs. Consequently the quantum state which encoded on the first pairs, decoded by the last pairs. In addition, it is observed that for particular values of $J$, QST does not happen. From time crystalline symmetry perspective, we have seen that entanglement oscillation illustrates the breaking of such a symmetry which originates in the nature of finite-size system. On the other hand, the interaction between a quantum system and its surroundings is inevitable. A quantum system would be influenced for example by phonons of the host crystal at $T\ne 0$. The system-environment interaction generates decoherence effects that can disrupt the quantum correlations
and can lead to other effects such as sudden death entanglement and sudden birth entanglement, as stated in Refs.\cite{motamedifar2019entanglement,yu2009sudden}.


\begin{thebibliography}{00}

\bibitem{chen2015expected}
X.~Chen, H.-M. Wang, D.-T. Ji, L.-Z. Mu and H.~Fan,
\newblock \emph{Expected number of quantum channels in quantum networks},
\newblock Scientific reports \textbf{5}, 12128 (2015).

\bibitem{bose2003quantum}
S.~Bose,
\newblock \emph{Quantum communication through an unmodulated spin chain},
\newblock Physical review letters \textbf{91}(20), 207901 (2003).

\bibitem{wang2001entanglement}
X.~Wang,
\newblock \emph{Entanglement in the quantum heisenberg xy model},
\newblock Physical Review A \textbf{64}(1), 012313 (2001).

\bibitem{feng2007entanglement}
R.~Feng-Hua and W.~Zhao-Ming,
\newblock \emph{Entanglement transfer via heisenberg interaction in a
	four-qubit system},
\newblock Communications in Theoretical Physics \textbf{47}(4), 621 (2007).

\bibitem{motamedifar2017dynamical}
M.~Motamedifar,
\newblock \emph{Dynamical pairwise entanglement and two-point correlations in
	the three-ligand spin-star structure},
\newblock Physica A: Statistical Mechanics and its Applications \textbf{483},
280 (2017).

\bibitem{motamedi2019exact}
S.~Shahsavari, M.~Motamedifar and H.~Safari,
\newblock \emph{Exact dynamics of concurrence-based entanglement in a system of
	four spin-1/2 particles on a triangular ladder structure},
\newblock Physica Scripta \textbf{95}(1), 015102 (2019).

\bibitem{wang2008quantum}
Z.-M. Wang, B.~Shao, P.~Chang and J.~Zou,
\newblock \emph{Quantum state transfer in a heisenberg xy chain with energy
	current},
\newblock Physica A: Statistical Mechanics and its Applications
\textbf{387}(8-9), 2197 (2008).

\bibitem{lyakhov2007role}
A.~Lyakhov, D.~Braun and C.~Bruder,
\newblock \emph{Role of interference in quantum state transfer through spin
	chains},
\newblock Physical Review A \textbf{76}(2), 022321 (2007).

\bibitem{boness2010doubly}
T.~Boness, K.~Kudo and T.~Monteiro,
\newblock \emph{Doubly excited ferromagnetic spin chain as a pair of coupled
	kicked rotors},
\newblock Physical Review E \textbf{81}(4), 046201 (2010).

\bibitem{oh2011heisenberg}
S.~Oh, L.-A. Wu, Y.-P. Shim, J.~Fei, M.~Friesen and X.~Hu,
\newblock \emph{Heisenberg spin bus as a robust transmission line for
	quantum-state transfer},
\newblock Physical Review A \textbf{84}(2), 022330 (2011).

\bibitem{giorgi2013quantum}
G.~L. Giorgi and T.~Busch,
\newblock \emph{Quantum state transfer in the presence of nonhomogeneous
	external potentials},
\newblock Physical Review A \textbf{88}(6), 062309 (2013).

\bibitem{lorenzo2015transfer}
S.~Lorenzo, T.~Apollaro, S.~Paganelli, G.~Palma and F.~Plastina,
\newblock \emph{Transfer of arbitrary two-qubit states via a spin chain},
\newblock Physical Review A \textbf{91}(4), 042321 (2015).

\bibitem{subrahmanyam2004entanglement}
V.~Subrahmanyam,
\newblock \emph{Entanglement dynamics and quantum-state transport in spin
	chains},
\newblock Physical Review A \textbf{69}(3), 034304 (2004).

\bibitem{apollaro2015many}
T.~Apollaro, S.~Lorenzo, A.~Sindona, S.~Paganelli, G.~Giorgi and F.~Plastina,
\newblock \emph{Many-qubit quantum state transfer via spin chains},
\newblock Physica Scripta \textbf{2015}(T165), 014036 (2015).

\bibitem{vieira2019robust}
R.~Vieira and G.~Rigolin,
\newblock \emph{Robust and efficient transport of two-qubit entanglement via
	disordered spin chains},
\newblock Quantum Information Processing \textbf{18}(5), 135 (2019).

\bibitem{apollaro2019spin}
T.~J. Apollaro, G.~M. Almeida, S.~Lorenzo, A.~Ferraro and S.~Paganelli,
\newblock \emph{Spin chains for two-qubit teleportation},
\newblock Physical Review A \textbf{100}(5), 052308 (2019).

\bibitem{shi2017robust}
X.~Shi, H.~Yuan, X.~Mao, Y.~Ma and H.~Zhao,
\newblock \emph{Robust quantum state transfer inspired by Dzyaloshinskii-Moriya
	interactions},
\newblock Physical Review A \textbf{95}(5), 052332 (2017).

\bibitem{mahmoudi2019effects}
M.~Mahmoudi,
\newblock \emph{The effects of Dzyaloshinskii-Moriya interaction on
	entanglement dynamics of a spin chain in a non-markovian regime},
\newblock Physica A: Statistical Mechanics and its Applications p. 123707
(2019).

\bibitem{messio2017chiral}
L.~Messio, S.~Bieri, C.~Lhuillier and B.~Bernu,
\newblock \emph{Chiral spin liquid on a kagome antiferromagnet induced by the
	Dzyaloshinskii-Moriya interaction},
\newblock Physical review letters \textbf{118}(26), 267201 (2017).

\bibitem{lee2018magnonic}
K.~H. Lee, S.~B. Chung, K.~Park and J.-G. Park,
\newblock \emph{Magnonic quantum spin hall state in the zigzag and stripe
	phases of the antiferromagnetic honeycomb lattice},
\newblock Physical Review B \textbf{97}(18), 180401 (2018).

\bibitem{farajolla}
T.~Farajollahpour and S.~Jafari,
\newblock \emph{Topological phase transition of the anisotropic x y model with
	Dzyaloshinskii-Moriya interaction},
\newblock Physical Review B \textbf{98}(8), 085136 (2018).

\bibitem{kato2019current}
N.~Kato, M.~Kawaguchi, Y.-C. Lau, T.~Kikuchi, Y.~Nakatani and M.~Hayashi,
\newblock \emph{Current-induced modulation of the interfacial
	Dzyaloshinskii-Moriya interaction},
\newblock Physical review letters \textbf{122}(25), 257205 (2019).

\bibitem{stagraczynski2017many}
S.~Stagraczy{\'n}ski, L.~Chotorlishvili, M.~Sch{\"u}ler, M.~Mierzejewski and
J.~Berakdar,
\newblock \emph{Many-body localization phase in a spin-driven chiral
	multiferroic chain},
\newblock Physical Review B \textbf{96}(5), 054440 (2017).

\bibitem{yokouchi2017electrical}
T.~Yokouchi, N.~Kanazawa, A.~Kikkawa, D.~Morikawa, K.~Shibata, T.~Arima,
Y.~Taguchi, F.~Kagawa and Y.~Tokura,
\newblock \emph{Electrical magnetochiral effect induced by chiral spin
	fluctuations},
\newblock Nature communications \textbf{8}(1), 1 (2017).

\bibitem{christandl2004perfect}
M.~Christandl, N.~Datta, A.~Ekert and A.~J. Landahl,
\newblock \emph{Perfect state transfer in quantum spin networks},
\newblock Physical review letters \textbf{92}(18), 187902 (2004).

\bibitem{guha2019allowed}
T.~Guha, M.~Alimuddin and P.~Parashar,
\newblock \emph{Allowed and forbidden bipartite correlations from thermal
	states},
\newblock Physical Review E \textbf{100}(1), 012147 (2019).

\bibitem{wilczek2012quantum}
F.~Wilczek,
\newblock \emph{Quantum time crystals},
\newblock Physical review letters \textbf{109}(16), 160401 (2012).

\bibitem{oberreiter2020stochastic}
L.~Oberreiter, U.~Seifert and A.~C. Barato,
\newblock \emph{Stochastic discrete time crystals: Entropy production and
	subharmonic synchronization},
\newblock arXiv preprint arXiv:2002.09078  (2020).

\bibitem{hill1997entanglement}
S.~Hill and W.~K. Wootters,
\newblock \emph{Entanglement of a pair of quantum bits},
\newblock Physical review letters \textbf{78}(26), 5022 (1997).

\bibitem{wootters1998entanglement}
W.~K. Wootters,
\newblock \emph{Entanglement of formation of an arbitrary state of two qubits},
\newblock Physical Review Letters \textbf{80}(10), 2245 (1998).

\bibitem{motamedifar2019entanglement}
M.~Motamedifar and M.~Golshani,
\newblock \emph{Entanglement dynamics in a spin star system coupled weakly to a
	bosonic bath},
\newblock Quantum Information Processing \textbf{18}(6), 181 (2019).

\bibitem{yu2009sudden}
T.~Yu and J.~Eberly,
\newblock \emph{Sudden death of entanglement},
\newblock Science \textbf{323}(5914), 598 (2009).



\end{thebibliography}


\end{document}